\documentclass[conference,twocolumn,oneside,10pt]{IEEEtran}
\IEEEoverridecommandlockouts

\usepackage{cite}

\ifCLASSINFOpdf
\usepackage[pdftex]{graphicx}
\else
\fi

\usepackage[cmex10]{amsmath}
\usepackage{amssymb,amsbsy,amsfonts,dsfont,amsthm,bm}

\usepackage{algorithmicx}

\usepackage{array}

\usepackage{mdwmath}
\usepackage{mdwtab}

\usepackage[]{caption}
\usepackage[caption=false,font=footnotesize]{subfig}

\usepackage{fixltx2e}

\usepackage{url}

\usepackage{latexsym,supertabular}
\usepackage{tikz}
\usetikzlibrary{shapes,arrows,shadows,automata,backgrounds,chains,matrix,patterns,petri,trees,calc}

\usepackage{balance}
\usepackage{multicol}
\usepackage{multirow}
\usepackage{booktabs}
\usepackage[capitalize]{cleveref}
\crefrangeformat{equation}{Equations~(#3#1#4) to~(#5#2#6)}

\newcommand{\bi}{\begin{itemize}}
\newcommand{\ei}{\end{itemize}}
\newcommand{\be}{\begin{eqnarray}}
\newcommand{\ee}{\end{eqnarray}}

\newcommand{\argmax}[1]{\underset{#1}{\text{arg max}}}
\newcommand{\argmin}[1]{\underset{#1}{\text{arg min}}}

\newcommand{\mx}[1]{\mathbf{\bm{#1}}} 
\newcommand{\vc}[1]{\mathbf{\bm{#1}}} 

\usepackage{eqparbox}

\usepackage{etoolbox}
\newtoggle{FigPosEndPaper}
\settoggle{FigPosEndPaper}{false} 

\newbool{EqOneColumn}
\setbool{EqOneColumn}{false} 

\usepackage{dblfloatfix}

\usepackage{float}
\usepackage[numbers,sort&compress,square,comma]{natbib}

\newcommand{\defoperator}{\stackrel{\text{\scriptsize def}}{=}}

\begin{document}

\title{Effects of Spatial Randomness on \\
	Locating a Point Source with Distributed Sensors
}


\author{
	\IEEEauthorblockN{Mohammad~Fanaei, 
	Matthew~C.~Valenti, 
	and
	Natalia~A.~Schmid
	}
	\IEEEauthorblockA{Lane Department of Computer Science and Electrical Engineering \\
		West Virginia University, Morgantown, WV, U.S.A. \\
		E-mail: mfanaei@mix.wvu.edu, 
			valenti@ieee.org, and natalia.schmid@mail.wvu.edu.}
}

\maketitle

\begin{abstract}
\boldmath
Most studies that consider the problem of estimating the location of a point source in wireless sensor networks assume that the source location is estimated by a set of spatially distributed sensors, whose locations are fixed. Motivated by the fact that the observation quality and performance of the localization algorithm depend on the location of the sensors, which could be randomly distributed, this paper investigates the performance of a recently proposed energy-based source-localization algorithm under the assumption that the sensors are positioned according to a uniform clustering process.
Practical considerations such as the existence and size of the exclusion zones around each sensor and the source will be studied. By introducing a novel performance measure called the estimation outage, it will be shown how parameters related to the network geometry such as the distance between the source and the closest sensor to it as well as the number of sensors within a region surrounding the source affect the localization performance.
\end{abstract}

\begin{IEEEkeywords}
Distributed source localization, distributed estimation, spatial randomness, binomial point process with repulsion, uniform clustering process, energy detector, fusion center, wireless sensor networks.
\end{IEEEkeywords}


\IEEEpeerreviewmaketitle

\section{Introduction}
\label{Sec:Intro}
%
%
%
The problem of energy-based source localization using a set of spatially distributed, randomly located, limited-power sensors forming a wireless sensor network (WSN) has recently attracted a lot of attention in the research community~\cite{OzdemirNiu09,NiuVarshney06,LiHu2003,Meesookho2008,Li2006,DardariConti2007}.
An effective source localization can be a first step in a broad range of other applications such as navigation, tracking, and geographic routing. In this context, the local sensors make noisy observations of the energy transmitted by, for example, an RF or acoustic source at their locations, process their noisy observations locally by, for instance, quantizing them, and send their processed data to a central entity in the network, known as the fusion center (FC), for further processing. The FC will then combine the received signals from local sensors, which are potentially corrupted by the communication channels between the sensors and itself, to estimate the location of the energy-transmitting source. As is common in the literature, it is reasonable to assume that the locations of the local sensors are known at the FC, which can be achieved using any form of cooperative localization schemes (e.g.,~\cite{WinConti2011,Patwari2005,ShenWin2010PartI,Wymeersch2009}).

The analyses and performance assessments in most of the works proposed in the literature for source localization can easily be generalized to a generic case in which the sensors are {\em randomly} located within the surveillance region covered by the network. Of course, the realization of the network geometry after its deployment should be known at the FC. However, the results of the performance analysis are usually presented for a fixed network topology such as a regular grid deployment~\cite{OzdemirNiu09} or for an average behavior of a number of random network realizations~\cite{Meesookho2008}. To the best of our knowledge, the effect of randomness of the sensor placement on the performance of source-localization schemes has been relatively unexplored, beyond analyzing the network's average behavior~\cite{Srinivasa2009,ZabiniGlobecom2011}. Srinivasa and Haenggi~\cite{Srinivasa2009} have considered the problem of distributed estimation of the path-loss exponent in an environment in which an RF signal is broadcasted, where the sensors are distributed according to a Poisson point process and sensor transmissions can interfere with each other.

The goal of this paper is to assess the performance of a typical source-localization scheme under different scenarios of random network realizations using numerical simulations. In other words, the question that we are trying to answer is as follows: Given a specific localization scheme, how does a randomly deployed WSN within a fixed surveillance region perform in terms of the localization accuracy? Note that answering this question gives significantly more insight into the design of a network than predicting only the {\em average} behavior of a randomly deployed system. Therefore, we are not proposing a new localization scheme, but rather we are applying concepts from stochastic geometry and point processes~\cite{Stoyan1996,Haenggi2012Book,WinPinto2009,Andrews2010} to investigate the performance of a refined and special version of a recently proposed source-localization algorithm~\cite{OzdemirNiu09}. A novel performance measure called the {\em localization outage} will be introduced to assess the performance of a typical localization algorithm. Numerical methods will be used to determine what parameters affect the performance of the given localization scheme when the sensors are placed according to a binomial point process with repulsion, which is also known as a uniform clustering process. The results of this analysis could be used to guide network deployment. If these guidelines are followed, a randomly formed network can be guaranteed (with some confidence) to achieve a minimum performance in terms of the localization accuracy.

The rest of this paper is organized as follows: Section~\ref{Sec:SystemModel} describes the system model considered in this paper. In Section~\ref{Sec:MLE_CRLB}, the source-localization scheme proposed in~\cite{OzdemirNiu09} is summarized and the Cram\'{e}r-Rao lower bound (CRLB) for the location estimator based on the binary quantized data at local sensors is derived. The effects of the random realization of the network geometry on the aforementioned localization scheme are shown through numerical simulations in Section~\ref{Sec:NumResultsMLECRLB}. Section~\ref{Sec:SpatialRandomness} introduces the concept of localization outage for a random network realization and discusses the effects of exclusion zones around local sensors on the performance of an arbitrary random realization of the network geometry. Finally, the paper is concluded in Section~\ref{Sec:Conclusions}.

\section{System Model}
\label{Sec:SystemModel}
\begin{figure}[!t]
	\centering
	\includegraphics[width=0.8\linewidth]{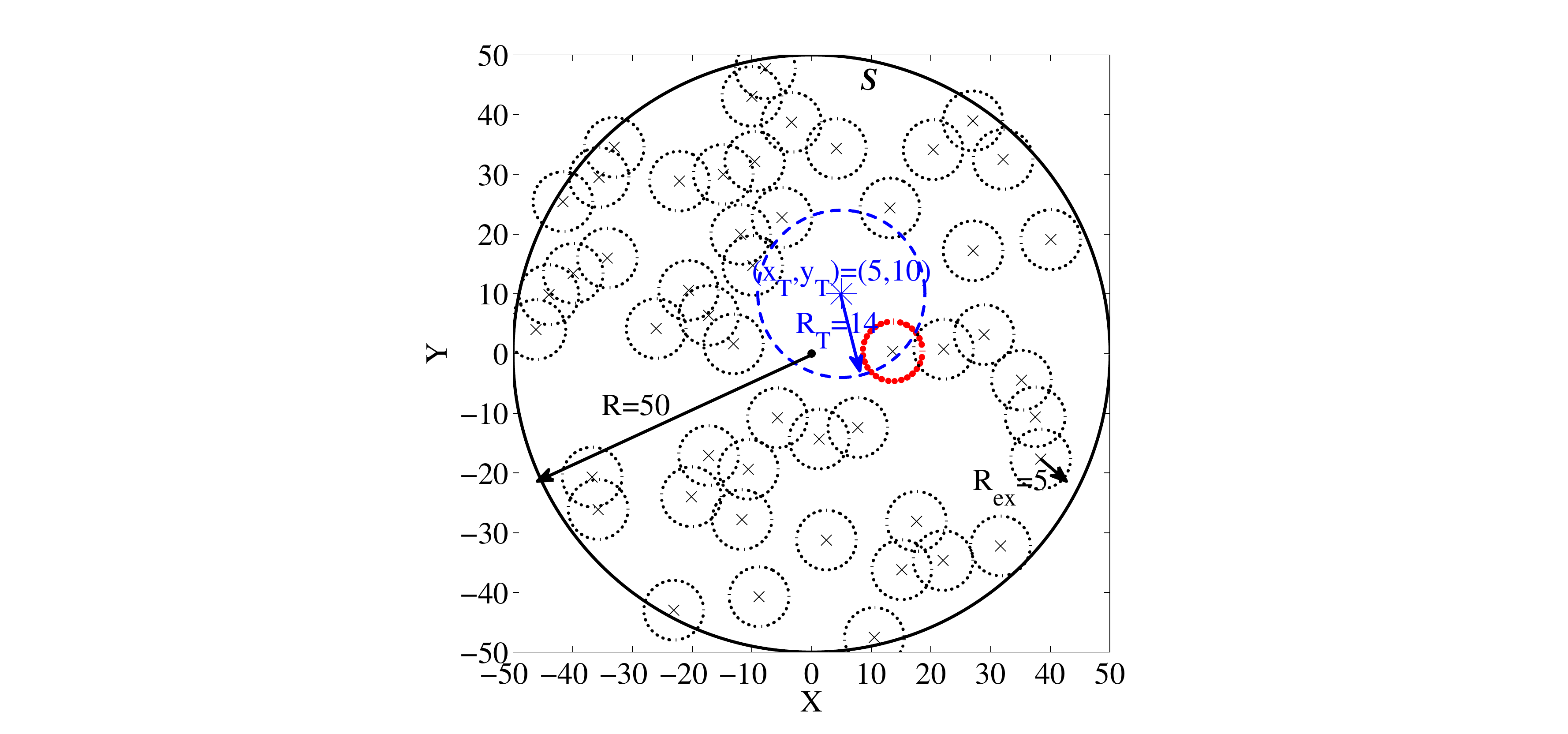}
	\caption{The network topology of an example WSN consisting of $K=50$ sensors denoted by `$\times$', whose objective is to localize a source target denoted by `{\color{blue} $\ast$}' and located at $\left( x_\mathsf{T}, y_\mathsf{T} \right) = \left( 5, 10 \right)$. The sensors are randomly placed in the circular surveillance region with radius $R=50$ and centered at the origin according to a uniform clustering process. Each sensor is surrounded by an exclusion zone with radius $R_{\mathsf{ex}} = 5$, shown by a dotted circle around the sensor. A dashed circle with radius $R_{\mathsf{T}} = 14$ is depicted around the source target, within which there is $K_{\mathsf{T}} = 1$ sensor enclosed.}
	\label{Fig:SampleNet1}
\end{figure}

Suppose that a WSN is composed of a fusion center (FC)
and $K$ sensors {\em arbitrarily} located in the two-dimensional space $\mathbb{R}^2$ within a circular surveillance region $\mathcal{S} \subseteq \mathbb{R}^2$ with radius $R$ and spatially distributed according to any point process.
Assume that a point source located at $\left( x_\mathsf{T}, y_\mathsf{T} \right) \in \mathcal{S}$ emits energy omni-directionally and that its power received by an arbitrary sensor $i$ located at $\left( x_i, y_i \right) \in \mathcal{S}$ is
\be \label{Eq:PoweDecayModel}
P_i
& = &
P_0 \left( \frac{d_0}{d_i} \right)^\alpha,
\quad
d_i \geq d_0,
\quad
i=1,2,\dotsc,K,
\ee
where $P_0$ is the received power from the source at the reference distance $d_0$, $\alpha$ is the power-decay exponent, and $d_i$ is the distance between the source and sensor $i$ defined as
\be
d_i
& = &
\sqrt{\left( x_\mathsf{T} - x_i \right)^2 + \left( y_\mathsf{T} - y_i \right)^2},
\quad
i=1,2,\dotsc,K.
\ee
An example of the random realization of such network topology is shown in Fig.~\ref{Fig:SampleNet1}, where $K=50$ sensors are randomly distributed in a circular region with radius $R=50$. Other parameters shown in the figure are introduced later in this paper. It should be mentioned that in addition to RF point sources, one of the most well-studied sources that satisfies the above power-decay model is the acoustic source, whose localization has widely been studied in the literature~\cite{LiHu2003}.

\begin{figure}[!t]
	\centering
	\includegraphics[width=1.04\linewidth]{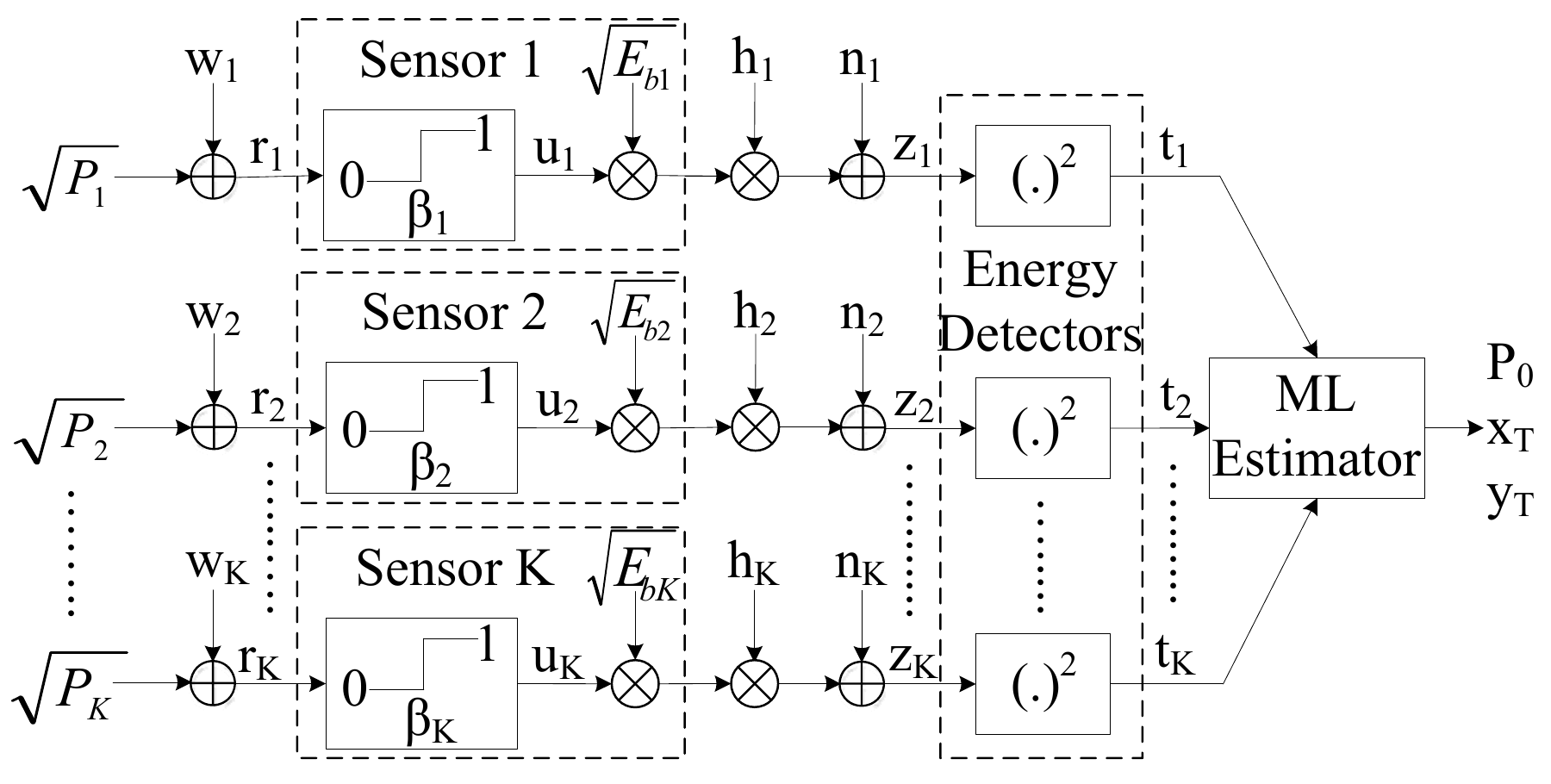}
	\caption{Functional system model of a WSN in which the FC localizes a source of energy.}
	\label{Fig:SysModel}
\end{figure}

Let $\vc{\theta} \defoperator \left[P_0, x_\mathsf{T}, y_\mathsf{T}\right]^T$ denote the vector of deterministic parameters associated with the source, where $\left[ \cdot \right]^T$ represents the transpose operation. The ultimate goal of the WSN is to estimate these parameters. More specifically, the focus of this paper is on the estimation of the source location. Figure~\ref{Fig:SysModel} shows the functional diagram of the WSN. Based on the above power-decay model, the received noisy signal at sensor $i$ is
\be \label{Eq:ObsModel}
r_i
& = &
\sqrt{P_i} + w_i,
\quad
i=1,2,\dotsc,K,
\ee
where $w_i$ is spatially independent additive white Gaussian noise (AWGN) with zero mean and variance $\sigma_i^2$, i.e., $w_i \sim \mathcal{N} \left( 0, \sigma_i^2 \right)$.
We define the {\em observation SNR} at sensor $i$ as $\mu_i \defoperator \frac{P_0}{\sigma_i^2}$, $i=1,2,\dotsc,K$. Upon observing the received noisy signal, each sensor uses a binary quantization scheme to quantize its local observation as
\be \label{Eq:LocalQuantization}
u_i
& = &
\begin{cases}
0, & \text{if } r_i < \beta_i \\
1, & \text{if } r_i \geq \beta_i
\end{cases},
\quad
i=1,2,\dotsc,K,
\ee
where $\beta_i$ is the binary quantization threshold at sensor $i$. Note that local sensors can process their noisy observations using various processing schemes. The simple binary quantization method considered as an example does not limit the generality of the following discussions and has only been used to emphasize the main objective of this paper, which is to study how spatial randomness affects the performance of a typical localization scheme.

Each sensor will use an on-off keying (OOK) scheme to send its quantized data to the FC through orthogonal channels corrupted by fading and AWGN. The received signal from sensor $i$ at the FC is
\be \label{Eq:ChannelModel}
z_i
& = &
h_i \sqrt{\mathcal{E}_{b_i}} \ u_i + n_i,
\quad
i=1,2,\dotsc,K,
\ee
where $\sqrt{\mathcal{E}_{b_i}} u_i$ is transmitted by the $i$th sensor,
$h_i$ is the multiplicative fading coefficient of the channel between sensor $i$ and the FC, and $n_i$ is the spatially independent, zero-mean, complex Gaussian random variable with
variance $\tau_i^2$, i.e., $n_i \sim \mathcal{CN} \left( 0, \tau_i^2 \right)$. In this paper, it is assumed that the channels between local sensors and the FC experience Rayleigh fading and therefore, the random variable $h_i$ is assumed to be spatially independent, zero-mean, complex Gaussian with unit power, i.e., $h_i \sim \mathcal{CN} \left( 0, 1 \right)$. It is also assumed that the FC does {\em not} have access to the instantaneous channel gains and that it only knows their distribution. We define the {\em channel SNR} for sensor $i$ as $\eta_i \defoperator \frac{\mathcal{E}_{b_i}}{\tau_i^2}$, $i=1,2,\dotsc,K$. Note that the above channel model implicitly assumes that the distance-dependent path-loss in the communication channels is fully compensated for all sensors using an appropriate power control scheme~\cite{Chiani2001}. Such power control makes the location of the FC irrelevant to the analysis. It should, however, be noted that sensor nodes that are farther from the FC will deplete their energy resources faster.

Upon receiving the signal from sensor $i$, the FC finds the energy of the received signal as $t_i = \lvert z_i \rvert^2$, $i = 1, 2, \dotsc, K$, where $\lvert \cdot \rvert$ denotes the absolute-value operation. Having access to $\vc{t} \defoperator \left[ t_1, t_2, \dotsc, t_K \right]^T$, the FC finds the maximum-likelihood (ML) estimate of the vector of unknown parameters $\vc{\theta}$ as explained in the following section.

\section{Derivation of ML Estimator and CRLB}
\label{Sec:MLE_CRLB}
As mentioned in the previous section, the sensors are {\em arbitrarily} located in the surveillance region $\mathcal{S}$. However, it is assumed that the FC knows their exact locations after the deployment of the WSN. This assumption could in practice be satisfied using any localization scheme~\cite{WinConti2011,Patwari2005,ShenWin2010PartI,Wymeersch2009}.
Note that the method and criteria for the localization of distributed sensors would in general be quite different from those of a single energy-emitting source, as considered in this paper.

Let $\Omega$ be a variable denoting a realization of the network geometry, when the WSN is deployed and the set of $\left\{ \left(x_i, y_i\right)\right\}_{i=1}^K$ (and consequently, the set of $\left\{ d_i\right\}_{i=1}^K$) is fixed. It is intuitive that the performance of any source-localization scheme, including the one studied in this paper, depends on the specific realization of the network geometry. The main goal of this paper is to study the effect of variable $\Omega$ as the network geometry on the performance of a source-localization scheme similar to the one proposed in~\cite{OzdemirNiu09}. In the rest of this section, the ML estimator and its corresponding CRLB proposed in~\cite{OzdemirNiu09} are summarized in order to assess the effect of network geometry on their performance.

\subsection{Derivation of ML Estimator}
\label{Subsec:MLDerivation}
Based on the observation model introduced in~\eqref{Eq:ObsModel} and the binary quantization rule specified in~\eqref{Eq:LocalQuantization}, the probability density function (pdf) of each sensor's quantized data parameterized by the vector of unknown parameters to be estimated, given a realization of the network geometry $\Omega$, can be found as
\ifbool{EqOneColumn}
{
	\be
	f_{U_i \vert \Omega}
	\left( u_i : \vc{\theta} \vert \Omega \right)
	& = &
	Q \left( \frac{\sqrt{P_i} - \beta_i}{\sigma_i} \right) \delta\left[ u_i \right]
	+
	Q \left( \frac{\beta_i - \sqrt{P_i} }{\sigma_i} \right) \delta\left[ u_i - 1 \right],
	\ee
}
{
	\begin{multline}
	f_{U_i \vert \Omega}
	\left( u_i : \vc{\theta} \vert \Omega \right)
	=
	Q \left( \frac{\sqrt{P_i} - \beta_i}{\sigma_i} \right) \delta\left[ u_i \right]
	\\
	+ Q \left( \frac{\beta_i - \sqrt{P_i} }{\sigma_i} \right) \delta\left[ u_i - 1 \right],
	\end{multline}
}
where $\delta\left[ \cdot \right]$ denotes the discrete Dirac delta function, and $Q(\cdot)$ is the complementary cumulative distribution function (CCDF) of the standard Gaussian random variable defined as
\be
Q(x)
& \defoperator &
\frac{1}{\sqrt{2 \pi}}
\int_x^\infty e^{-\frac{t^2}{2}} \, \mathrm{d}t.
\ee

Based on the channel model introduced in~\eqref{Eq:ChannelModel}, given any binary sensor decision $u_i$, the signal received from sensor $i$ at the FC is a complex Gaussian random variable with zero mean and variance $\mathcal{E}_{b_i} u_i^2 + \tau_i^2$, i.e., $z_i \vert u_i \sim \mathcal{CN} \left( 0, \mathcal{E}_{b_i} u_i^2 + \tau_i^2 \right)$. Note that the channel fading coefficient and the channel AWGN are assumed to be independent. Based on this result, the energy of the received signal from sensor $i$ at the FC, given the sensor's binary decision, is exponentially distributed with parameter $\lambda_i \defoperator \frac{1}{\mathcal{E}_{b_i} u_i^2 + \tau_i^2}$, i.e., $t_i \vert u_i \sim \mathcal{E} \left( \frac{1}{\mathcal{E}_{b_i} u_i^2 + \tau_i^2} \right)$. Therefore, given a realization of the network geometry $\Omega$, the joint pdf of the vector of received energies from different sensors at the FC, parameterized by the vector of unknown parameters to be estimated can be written as
\vspace{-1pt}
\be \label{Eq:EnergyJointPdf}
f_{\vc{T} \vert \Omega} \left(\vc{t} : \vc{\theta} \vert \Omega\right)
& = &
\prod_{i=1}^K f_{T_i \vert \Omega} \left(t_i : \vc{\theta} \vert \Omega\right),
\ee
where
\be \label{Eq:EnergyMarginalPdf}
f_{T_i \vert \Omega} \left(t_i : \vc{\theta} \vert \Omega\right)
& = &
\int f_{T_i \vert U_i} \left(t_i \vert u_i\right) f_{U_i \vert \Omega} \left( u_i : \vc{\theta} \vert \Omega \right) \, \mathrm{d}u_i \notag \\
& \overset{(a)}{=} &
\frac{1}{\tau_i^2} e^{- \frac{t_i}{\tau_i^2}} Q \left( \frac{\sqrt{P_i} - \beta_i}{\sigma_i} \right) \\
& & \hspace{0.5cm} +
\frac{1}{\mathcal{E}_{b_i} + \tau_i^2} e^{- \frac{t_i}{\mathcal{E}_{b_i} + \tau_i^2}} Q \left( \frac{\beta_i - \sqrt{P_i}}{\sigma_i} \right) \notag
\ee
where $(a)$ is based on the sifting property of the Dirac delta function. It is well known that the ML estimate of the vector of unknown parameters at the FC using the vector of the received energies from local sensors can be found as~\cite[Chapter 7]{Kay93}
\be
\hat{\vc{\theta}}_\Omega
& = &
\argmax{\vc{\theta}} \; \ln \left( f_{\vc{T} \vert \Omega} \left(\vc{t} : \vc{\theta} \vert \Omega\right) \right),
\ee
where the subscript $\Omega$ for the ML estimator signifies that it depends on the realization of the network geometry.

\subsection{Derivation of CRLB}
\label{Subsec:CRLBDerivation}
The performance of any estimator can be quantified by its variance. The Cram\'{e}r-Rao lower bound (CRLB) expresses a lower bound on the variance of any {\em unbiased} estimator $\hat{\vc{\theta}}_\Omega$ as~\cite[Chapter 3]{Kay93}
\be \label{Eq:CRLBDef}
\mathbb{E}
\left[
\left( \hat{\vc{\theta}}_\Omega - \vc{\theta} \right) \left( \hat{\vc{\theta}}_\Omega - \vc{\theta} \right)^T
\right]
& \succeq &
\mx{\mathcal{I}}_\Omega^{-1} \left( \vc{\theta} \right),
\ee
where $\mathbb{E} \left[ \cdot \right]$ represents the expectation operation with respect to the joint pdf of the vector of received energies from different sensors at the FC, $\mx{\Phi} \succeq \mx{\Lambda}$ means that the matrix $\mx{\Phi} - \mx{\Lambda}$ is positive semi-definite, and $\mx{\mathcal{I}}_\Omega \left( \vc{\theta} \right)$ denotes the Fisher information matrix (FIM) for the given realization of the network geometry $\Omega$, whose element in row $m$ and column $n$ is defined as
\be
\left[ \mx{\mathcal{I}}_\Omega \left( \vc{\theta} \right) \right]_{m,n}
& = &
- \mathbb{E}
\left[
\frac{\partial^2 \ln \left( f_{\vc{T} \vert \Omega} \left(\vc{t} : \vc{\theta} \vert \Omega\right) \right)}{\partial \theta_m \partial \theta_n}
\right].
\ee
Based on the joint pdf of the vector of received energies from different sensors at the FC defined in~\eqref{Eq:EnergyJointPdf}--\eqref{Eq:EnergyMarginalPdf}, the FIM for the given observation and channel models and given a realization of the network geometry $\Omega$ can be found as follows~\cite{OzdemirNiu09}:
\be \label{Eq:FIM_Def}
\mx{\mathcal{I}}_\Omega \left( \vc{\theta} \right)
=
\sum_{i = 1}^K
\frac{\mx{\mathcal{G}}_i P_i}{8 \pi  \sigma_i^2 P_0}
e^{- \frac{\left( \sqrt{P_i} - \beta_i \right)^2}{\sigma_i^2}}
\int_0^\infty
\frac{g\left(t_i \right) \, \mathrm{d} t_i}{f_{T_i \vert \Omega} \left(t_i : \vc{\theta} \vert \Omega\right)}
\ee
where $g\left(t_i \right)$ is defined as
\be
g\left(t_i \right)
& \defoperator &
\left(
\frac{1}{\mathcal{E}_{b_i} + \tau_i^2} e^{- \frac{t_i}{\mathcal{E}_{b_i} + \tau_i^2}}
-
\frac{1}{\tau_i^2} e^{- \frac{t_i}{\tau_i^2}}
\right)^2,
\ee
$f_{T_i \vert \Omega} \left(t_i : \vc{\theta} \vert \Omega\right)$ in found in~\eqref{Eq:EnergyMarginalPdf}, and $\mx{\mathcal{G}}_i \defoperator \mx{\mathcal{G}}_{i,\Omega} \left( \vc{\theta} \right)$ is a symmetric 3-by-3 matrix, whose elements are defined as follows:
\be \label{Eq:GMatDef}
& & \left[ \mx{\mathcal{G}}_{i,\Omega} \left( \vc{\theta} \right) \right]_{1,1} = \frac{1}{P_0} \qquad
\left[ \mx{\mathcal{G}}_{i,\Omega} \left( \vc{\theta} \right) \right]_{2,2} = \frac{P_0 \alpha^2}{d_i^4} \left( x_\mathsf{T} - x_i \right)^2 \notag \\
& & \left[ \mx{\mathcal{G}}_{i,\Omega} \left( \vc{\theta} \right) \right]_{1,2} = \left[ \mx{\mathcal{G}}_{i,\Omega} \left( \vc{\theta} \right) \right]_{2,1} = \frac{\alpha }{d_i^2} \left( x_i - x_\mathsf{T} \right) \notag \\
& & \left[ \mx{\mathcal{G}}_{i,\Omega} \left( \vc{\theta} \right) \right]_{1,3} = \left[ \mx{\mathcal{G}}_{i,\Omega} \left( \vc{\theta} \right) \right]_{3,1} = \frac{\alpha}{d_i^2} \left( y_i - y_\mathsf{T} \right) \\
& & \left[ \mx{\mathcal{G}}_{i,\Omega} \left( \vc{\theta} \right) \right]_{2,3} = \left[ \mx{\mathcal{G}}_{i,\Omega} \left( \vc{\theta} \right) \right]_{3,2} = \frac{P_0 \alpha^2}{d_i^4} \left( x_\mathsf{T} - x_i \right) \left( y_\mathsf{T} - y_i \right) \notag \\
& & \left[ \mx{\mathcal{G}}_{i,\Omega} \left( \vc{\theta} \right) \right]_{3,3} = \frac{P_0 \alpha^2}{d_i^4} \left( y_\mathsf{T} - y_i \right)^2 \notag
\ee
Note that the matrix $\mx{\mathcal{G}}_{i,\Omega} \left( \vc{\theta} \right)$ and consequently, the FIM and CRLB depend on the realization of the network geometry.

\subsection{Performance-Assessment Metric for Localization Schemes}
\label{Subsec:GLEMonteCarlo}
One of the main measures used to assess the performance of any source-localization scheme is the {\em geometric location-estimation error} (GLE) defined as~\cite{Li2006}
\be \label{Eq:GLEDef}
\mathsf{GLE}_\Omega
& \defoperator &
\sqrt{\left( \hat{x}_{\mathsf{T},\Omega} - x_\mathsf{T} \right)^2 + \left( \hat{y}_{\mathsf{T},\Omega} - y_\mathsf{T} \right)^2},
\ee
where the subscript $\Omega$ signifies that the GLE depends on the realization of the network geometry. Note that given a specific realization of the network geometry $\Omega$, the following lower bound can be established on the mean squared GLE using the CRLB as defined in~\eqref{Eq:CRLBDef}:
\be \label{Eq:MS_GLE_Def}
\overline{\mathsf{SGLE}}_\Omega
& \defoperator &
\mathbb{E} \left[ \mathsf{GLE}_\Omega^2 \right]
=
\mathbb{E} \left[
\left( \hat{x}_{\mathsf{T},\Omega} - x_\mathsf{T} \right)^2 + \left( \hat{y}_{\mathsf{T},\Omega} - y_\mathsf{T} \right)^2
\right] \notag \\
& \geq &
\left[ \mx{\mathcal{I}}_\Omega^{-1} \left( \vc{\theta} \right) \right]_{2,2}
+
\left[ \mx{\mathcal{I}}_\Omega^{-1} \left( \vc{\theta} \right) \right]_{3,3},
\ee
where $\overline{\mathsf{SGLE}}_\Omega$ denotes the mean squared GLE, given a specific realization of the network geometry $\Omega$, and the expectation operation is calculated with respect to the distributions of the observation noise, channel fading coefficients, and channel noise.

Since there is no closed-form equation for finding $\overline{\mathsf{SGLE}}_\Omega$, we resort to a Monte-Carlo approach for its calculation as follows. For a fixed arbitrary realization of the network geometry $\Omega$, the set of sensors' locations $\left\{ \left(x_i, y_i\right)\right\}_{i=1}^K$ and consequently, the sets of their distances to the source target $\left\{ d_i\right\}_{i=1}^K$ and the received power from the source at their locations $\left\{ P_i\right\}_{i=1}^K$ (defined in~\eqref{Eq:PoweDecayModel}) are fixed. In order to find the empirical mean squared GLE, $N_{\mathsf{MC}}$ Monte-Carlo trials are performed for the given network geometry by generating random observation noises, channel fading coefficients, and channel noises based on their respective distributions introduced in Section~\ref{Sec:SystemModel}. The empirical mean squared GLE for the given network realization can then be found as
\be
\overline{\mathsf{SGLE}_\Omega}
& = &
\frac{1}{N_{\mathsf{MC}}}
\sum_{m = 1}^{N_{\mathsf{MC}}} \mathsf{SGLE}_\Omega^{(m)} \\
& = &
\frac{1}{N_{\mathsf{MC}}}
\sum_{m = 1}^{N_{\mathsf{MC}}}
\left( \hat{x}_{\mathsf{T},\Omega}^{(m)} - x_\mathsf{T} \right)^2 + \left( \hat{y}_{\mathsf{T},\Omega}^{(m)} - y_\mathsf{T} \right)^2, \notag
\ee
where $\mathsf{SGLE}_\Omega \defoperator \mathsf{GLE}_\Omega^2$ and $\mathsf{GLE}_\Omega$ is defined in~\eqref{Eq:GLEDef}, and the superscript $(m)$ denotes the result obtained in the $m$th Monte-Carlo trial.

\subsection{Derivation of Optimal Local Quantization Thresholds}
\label{Subsec:QuantizationThreshold}
Note that the performance of both empirical mean squared GLE and its corresponding CRLB for any fixed network realization is a function of the local sensors' binary quantization thresholds. In this paper, the optimal set of local quantization thresholds are found based on the approach proposed in~\cite{NiuVarshney06}. According to this method, since the main focus of this paper is the accurate localization of the source target and {\em not} so much the accurate estimation of $P_0$ as the received power from the source at the reference distance $d_0$, the binary quantization thresholds are found such that the CRLB on the mean squared GLE as defined in~\eqref{Eq:MS_GLE_Def} is minimized. In other words, the optimal set of binary quantization thresholds for the optimal source-localization scheme can be found as~\cite{NiuVarshney06}
\be
\left\{ \beta_i^{\mathsf{OPT}} \right\}_{i = 1}^K
=
\argmin{\left\{ \beta_i \right\}_{i = 1}^K}
\left(
\left[ \mx{\mathcal{I}}_\Omega^{-1} \left( \vc{\theta} \right) \right]_{2,2}
+
\left[ \mx{\mathcal{I}}_\Omega^{-1} \left( \vc{\theta} \right) \right]_{3,3}
\right),
\ee
where the conditional FIM $\mx{\mathcal{I}}_\Omega \left( \vc{\theta} \right)$, given the current network realization $\Omega$ is found using~\eqref{Eq:FIM_Def}--\eqref{Eq:GMatDef}.

\section{Numerical Performance Assessment}
\label{Sec:NumResultsMLECRLB}
Ozdemir~et~al.~\cite{OzdemirNiu09} have reported the performance of their proposed source-localization scheme summarized in Subsections~\ref{Subsec:MLDerivation} and \ref{Subsec:CRLBDerivation} for a WSN deployed in a regular grid configuration. As mentioned previously, the performance of the source-localization method is heavily affected by the realization of the network geometry. In order to observe this dependence, suppose that a WSN consisting of $K=50$ sensors is randomly deployed to estimate the location and parameter $P_0$ of a source target located at $\left( x_\mathsf{T}, y_\mathsf{T} \right) = \left( 5, 10 \right)$, for which $P_0 = 10,000$, $d_0 = 1$, and the power-decay exponent is $\alpha = 2$. The sensors are randomly placed in the circular surveillance region with radius $R=50$ and centered at the origin according to a {\em uniform clustering process}. The local observation noises are assumed to be identically distributed with the same variance $\sigma^2 \equiv \sigma_i^2 = \frac{P_0}{\mu}$, where $\mu \equiv \mu_i$ is the common observation SNR. Similarly, the local channel noises are assumed to be identically distributed with the same variance $\tau^2 \equiv \tau_i^2 = \frac{\mathcal{E}_{b}}{\eta}$, where $\mathcal{E}_{b} \equiv \mathcal{E}_{b_i} = 1 \ \mathrm{dB}$ is the common transmit energy when $u_i = 1$ is sent, and $\eta \equiv \eta_i$ is the common channel SNR. Due to the homogeneous nature of the network, all of the binary quantization thresholds are assumed to be identical to $\beta \equiv \beta_i$. The results have been found by averaging over $N_{\mathsf{MC}} = 10,000$ Monte-Carlo trials as explained in Subsection~\ref{Subsec:GLEMonteCarlo}.

\begin{figure}[!t]
	\centering
	{\includegraphics[width=1.02\linewidth]{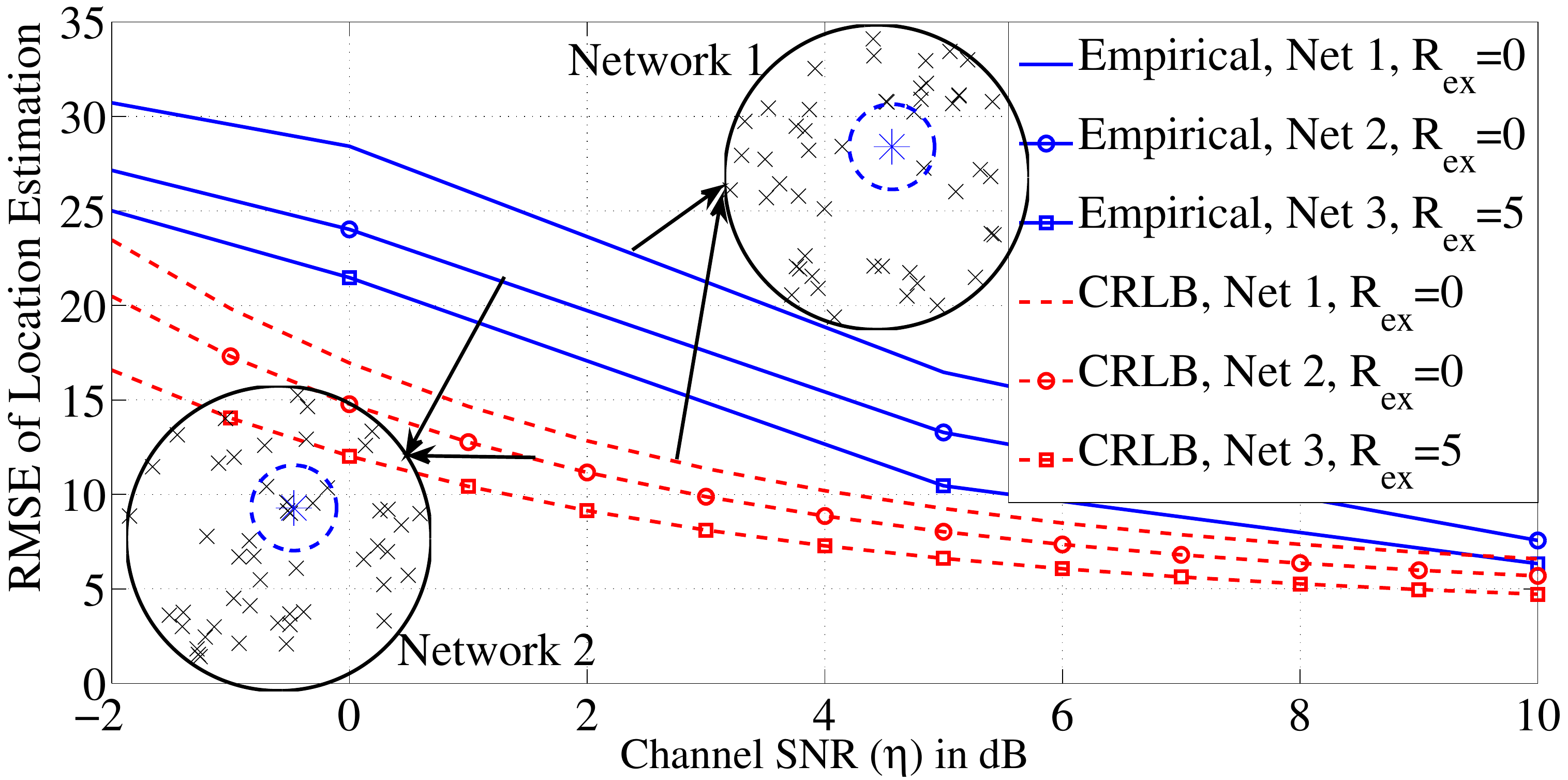}}
	\caption{Empirical RMSE of the source-location estimation, shown by solid lines, and its corresponding CRLB, shown by dashed lines, vs channel SNR in dB for three different random realizations of the network geometry, when the observation SNR is $\mu = 40 \ \mathrm{dB}$.}
	\label{Fig:OSNR40_FixedNet_GLE}
\end{figure}

Figure~\ref{Fig:OSNR40_FixedNet_GLE} shows the empirical root mean-squared error (RMSE) for the source-location estimation, plotted by solid lines, and its corresponding CRLB, plotted by dashed lines, as functions of the channel SNR $\eta$ for three different random network realizations, when the observation SNR is fixed at $\mu = 40 \ \mathrm{dB}$. Details of generating each random network geometry are explained in the next section. The first and second network realizations corresponding to the curves without marker and with the circle marker, respectively, are shown in the corners of Fig.~\ref{Fig:OSNR40_FixedNet_GLE}, where sensors are denoted by `$\times$'. In these two network realizations, there is no exclusion zone considered around the sensors (i.e., $R_{\mathsf{ex}} = 0$) and therefore, they can be placed very close to each other. The network realization corresponding to the curves shown by the square marker `$\Box$' is depicted in Fig.~\ref{Fig:SampleNet1}. In this case, each sensor is surrounded by an exclusion zone with radius $R_{\mathsf{ex}} = 5$ and therefore, all of the sensors will be apart from each other by at least 5 units of length. It can be seen in this figure that the performance of the source-localization scheme highly depends on the realization of the network geometry. It also shows that as the channel SNR increases, the error in the localization decreases and gets closer to its CRLB, as expected. Similar results could be found by considering the localization performance as a function of the observation SNR.

A close look at all network configurations shown in Fig.~\ref{Fig:SampleNet1} and the corners of Fig.~\ref{Fig:OSNR40_FixedNet_GLE} reveals that a circle with radius $R_{\mathsf{T}} = 14$ is centered at the source target and shown by a dashed line. In Network 1 shown at the top of Fig.~\ref{Fig:OSNR40_FixedNet_GLE}, there are only two sensors located within this region surrounding the target, while in Network 2 shown at the bottom of this figure, there are six such sensors in the same vicinity of the target. This difference would partially explain why the performance of the localization scheme using the two different network realizations is completely different. When there are more sensors within the immediate vicinity of the target, the localization error will be lower since the observations are of higher quality. This point will further be discussed with more details in Subsection~\ref{Subsec:SensorDistance2Target}.

\section{Spatial Dependence of Source Localization}
\label{Sec:SpatialRandomness}
In this section, we study the effects of spatial randomness, i.e., random realization of the network geometry, on the performance of the source-localization scheme proposed in~\cite{OzdemirNiu09} and summarized in Section~\ref{Sec:MLE_CRLB}, through a numerical Monte-Carlo approach. Note that the performance evaluations presented here can easily be extended to any other source-localization method.

Let the {\em localization outage event} in the space of random realizations of network geometry be defined as
\be
\mathcal{A} \left(\gamma\right)
& \defoperator &
\left\{ \Omega : \overline{\mathsf{SGLE}}_\Omega > \gamma^2 \right\}.
\ee
Based on the above definition, a {\em localization outage} occurs when the root mean squared distance between the estimated location of the source $\left( \hat{x}_\mathsf{T}, \hat{y}_\mathsf{T} \right)$ and its true value $\left( x_\mathsf{T}, y_\mathsf{T} \right)$ exceeds a prespecified threshold $\gamma$. In other words, a realization of the network geometry $\Omega$ is said to be in outage if {\em on average}, the source-location estimation using that network deployment produces an error beyond an acceptable threshold $\gamma$.

It can be observed that the localization outage is a random variable depending on the distribution of the network geometry. In order to assess the dependence of the localization outage on the realization of the network geometry, we will use the complementary cumulative distribution function (CCDF) of the random variable $\overline{\mathsf{SGLE}}_\Omega$ defined as
\be \label{Eq:CCDF_Definition}
\overline{F}_{\overline{\mathsf{SGLE}}} \left( \gamma \right)
& \defoperator &
\mathbb{P} \left[ \overline{\mathsf{SGLE}}_\Omega > \gamma^2 \right],
\ee
where the right-hand side of the equation is the probability that an arbitrary network geometry is in outage, as defined above.

In the following discussions, the performance assessments will be based on a Monte-Carlo approach with 500 simulation trials ran as follows. In each simulation trial, a realization of the network geometry is obtained by randomly placing $K$ sensors within the circular region of radius $R$. There could be an exclusion zone around each sensor within which no other sensors can be placed. The sensors are located within the surveillance region successively according to a {\em uniform clustering process} as follows. A pair of independent random variables $\left(x_i,y_i\right)$, $i = 1,2,\dotsc,K$, is selected from a uniform distribution over $\left[-R,R\right]$. If the sensor falls outside of the circular disk of radius $R$, i.e., $x_i^2 + y_i^2 > R^2$, this process is repeated until the sensor falls inside the surveillance region. If an exclusion zone of radius $R_{\mathsf{ex}}$ is considered around each sensor, the distances between the $i$th sensor and all $i-1$ other sensors are found, and the above process of assigning new random location to the $i$th sensor is repeated as many times as necessary until the sensor is located outside of the exclusion zones of all other previously located sensors in the network.

In the next step, the empirical mean squared GLE (i.e., $\overline{\mathsf{SGLE}}_\Omega$) is found for a fixed random realization of the network geometry $\Omega$ using the Monte-Carlo approach described in Subsection~\ref{Subsec:GLEMonteCarlo} with $N_{\mathsf{MC}} = 1,000$ trials per network realization. The CRLB on the mean squared GLE is found {\em only once} for each network realization as defined in~\eqref{Eq:MS_GLE_Def}.
The optimal local binary quantization threshold is fixed and found only once for any network realization, using the approach discussed in Subsection~\ref{Subsec:QuantizationThreshold}. All of the simulation parameters are exactly the same as those summarized in Section~\ref{Sec:NumResultsMLECRLB}. The observation SNR and channel SNR are fixed at $\mu = 40 \ \mathrm{dB}$ and $\eta = 0 \ \mathrm{dB}$, respectively.

\subsection{Effect of Sensor Exclusion Zones on Source Localization}
\begin{figure}[!t]
	\centering
	{\includegraphics[width=1.04\linewidth]{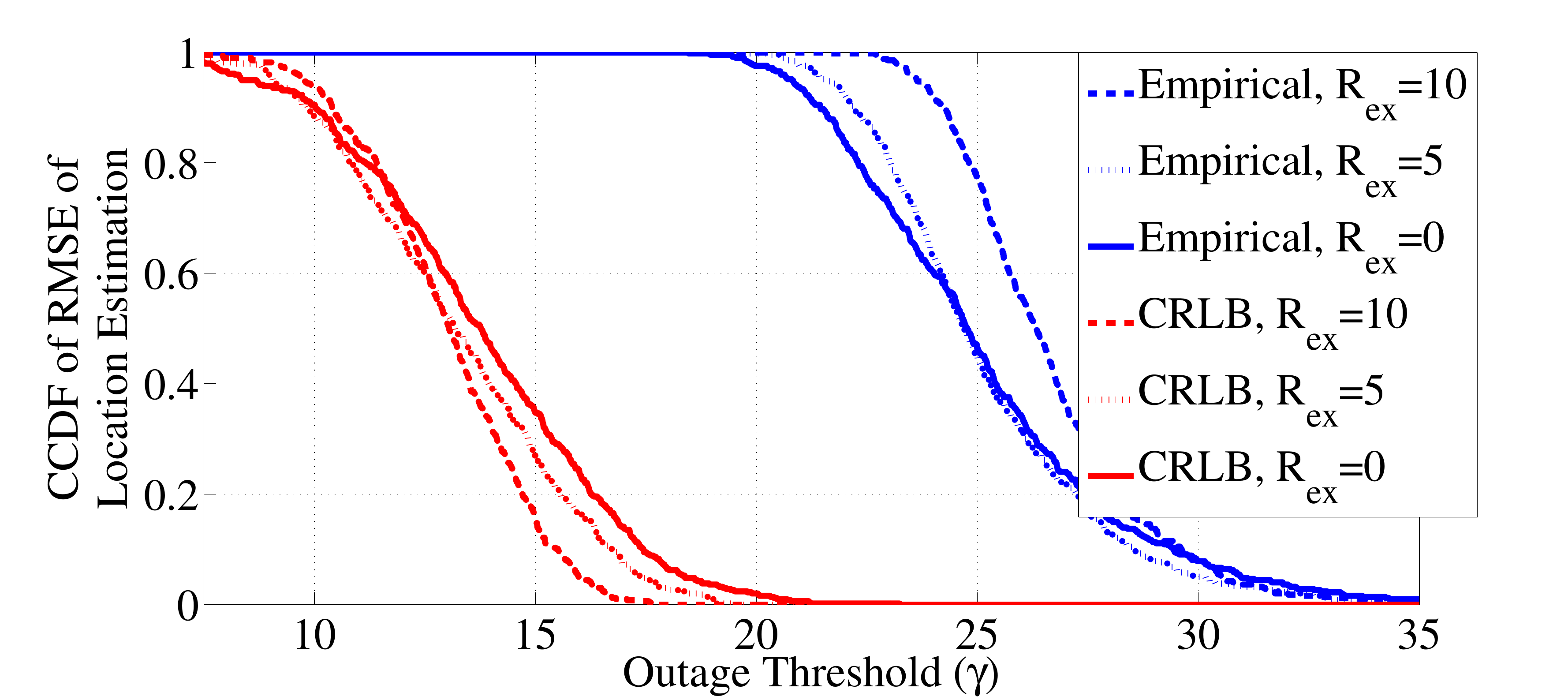}}
	\caption{CCDF of the empirical RMSE of the source-location estimation and its corresponding CRLB vs the outage threshold $\gamma$ for different settings of the network geometry.
	The observation SNR and channel SNR are fixed at $\mu = 40 \ \mathrm{dB}$ and $\eta = 0 \ \mathrm{dB}$, respectively.}
	\label{Fig:CCDF_GLE_SEx_RandNet}
\end{figure}

One of the parameters affecting the performance of any source-localization scheme, which is based on the assumption of random sensor placement, is the {\em exclusion zone} around each sensor. The exclusion zone is a circular disk around each sensor within which no other sensors can be placed. It could be the result of a physical limitation that does not allow such proximity of two sensors or it could be controlled by the network administrator during the network deployment in order to guarantee a proper coverage of the surveillance region. Figure~\ref{Fig:CCDF_GLE_SEx_RandNet} depicts the CCDF of the empirical RMSE of the source-location estimation, as defined by~\eqref{Eq:CCDF_Definition}, and its corresponding CRLB as functions of the outage threshold $\gamma$ for different values of the radius of sensor exclusion zones $R_{\mathsf{ex}}$. The results were obtained using 500 Monte-Carlo trials for generating random network realizations as described in the previous subsection.

As it can be seen from Fig.~\ref{Fig:CCDF_GLE_SEx_RandNet}, the probability of an empirical localization outage increases as the radius of the sensor exclusion zones increases. In other words, as the exclusion zone around each sensor expands, the probability that the average GLE of an arbitrary random network deployment exceeds a prescribed threshold increases, due to the fact that the expansion of the exclusion zones around sensors results in them being located farther apart. Therefore, the number of sensors that can be located close to a target decreases on average. This will result in a lower number of strong local measurements, which in turn decreases the quality of the data available at the FC as more sensors are likely to have sent zeros. It should be mentioned that for lower probabilities of localization outage, i.e., higher values of outage threshold $\gamma$, the exclusion zones around sensors do not have much effect as almost any network realization can on average satisfy the required accuracy of location estimation. The same argument applies to the CCDF for the CRLB values on the root mean squared location estimation error.

\subsection{Effect of the Closest Sensors to Source on Localization}
\label{Subsec:SensorDistance2Target}
It is intuitive that the performance of any source-localization scheme depends mainly on the observation and channel qualities of the closest sensors to the target. In order to investigate this effect, consider a scenario in which there is no exclusion zone around the sensors, i.e., $R_{\mathsf{ex}} = 0$. Note that similar results and discussions could be found for the network realizations with an arbitrary exclusion zone around sensors. Let $R_{\mathsf{T}}$ denote the radius of a circular region around the target within which we assume the most important sensors to the performance of the source-localization scheme are located. Let $K_{\mathsf{T}}$ denote the number of sensors located within this region. In the network realization depicted in Fig.~\ref{Fig:SampleNet1}, the region around the target is shown by a dashed line as a circle with radius $R_{\mathsf{T}} = 14$, and the number of sensors within this region is $K_{\mathsf{T}} = 1$. Note that in general, $0 \leq R_{\mathsf{T}} \leq 2R$ and $0 \leq K_{\mathsf{T}} \leq K$, where $R$ is the radius of the surveillance region. Figure~\ref{Fig:CCDF_GLE_NoSEx_KT_RandNet} depicts the CCDF of the empirical RMSE of the source-location estimation, as defined by~\eqref{Eq:CCDF_Definition}, and its corresponding CRLB as functions of the outage threshold $\gamma$ for different values of $R_{\mathsf{T}}$ and $K_{\mathsf{T}}$, when there is no exclusion zone around the sensors. The results were obtained in a similar way to the procedure explained at the beginning of this section.

As it can be seen in Fig.~\ref{Fig:CCDF_GLE_NoSEx_KT_RandNet}, for a given $R_{\mathsf{T}}$, the probability of localization outage decreases as $K_{\mathsf{T}}$ increases. In other words, if in the random realizations of the network geometry, the number of sensors located within a fixed radius around the target increases, the probability that an arbitrary network deployment is in outage drastically decreases. In a similar discussion, for a given $K_{\mathsf{T}}$, the probability of outage decreases as the radius $R_{\mathsf{T}}$ decreases. In other words, if we need to expand the region around the target to have a specific, fixed number of sensors located close to it, the probability of outage increases as the region expands. Figure~\ref{Fig:CCDF_GLE_NoSEx_KT_RandNet} shows that the effect of increasing $K_{\mathsf{T}}$ for a fixed $R_{\mathsf{T}}$ is always noticeable, while the effect of decreasing $R_{\mathsf{T}}$ for a fixed $K_{\mathsf{T}}$ is more noticeable when the number of sensors considered within the neighborhood of the target is larger. The important implication of this discussion in practical network design is that the density of the randomly deployed network should be above a threshold to guarantee that the sensors are so closely located that if the target location is anywhere within the surveillance region, there are enough number of sensors in its proximity.
{\setlength{\belowcaptionskip}{-8pt}
\begin{figure}[!t]
	\centering
	{\includegraphics[width=0.93\linewidth]{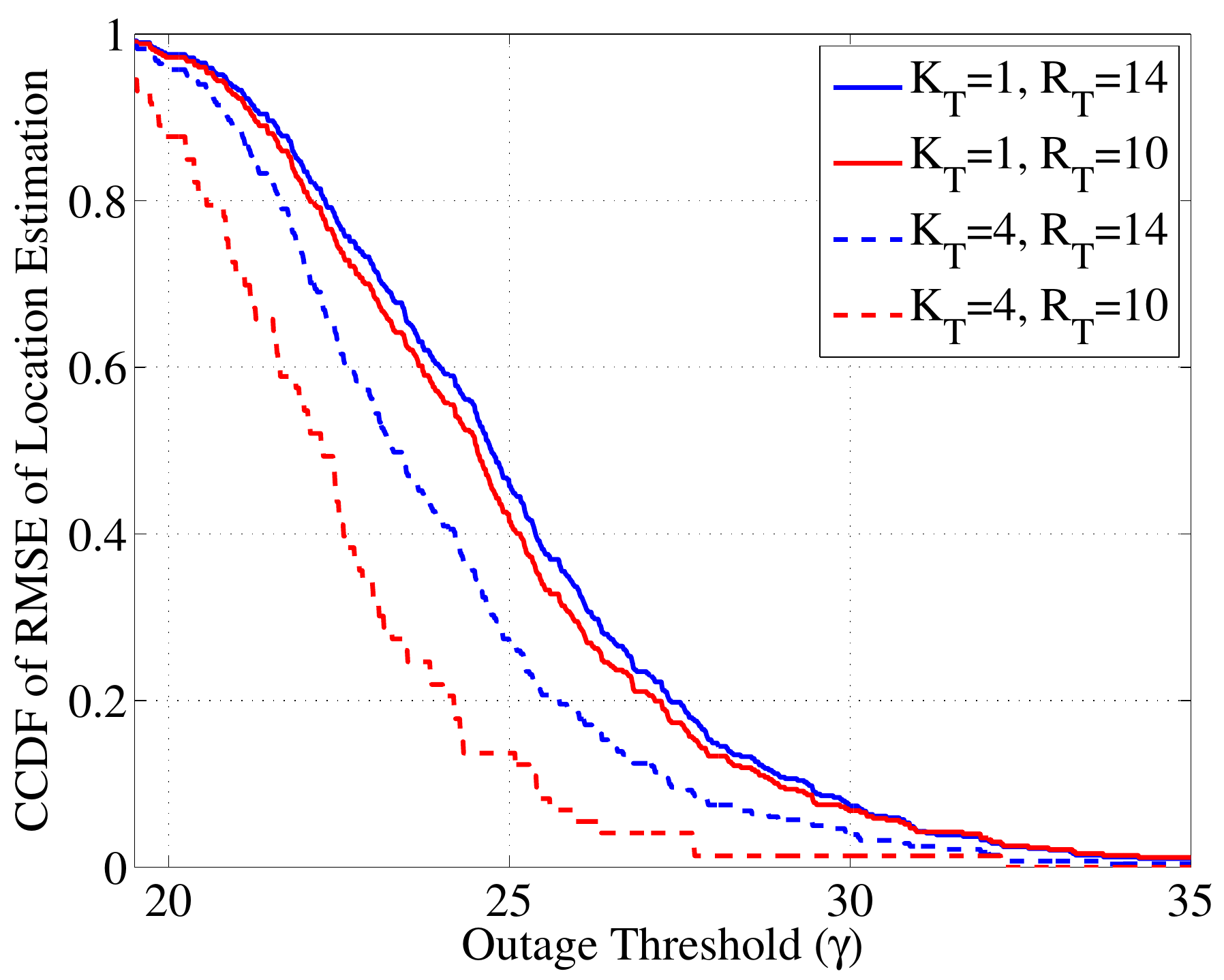}}
	\caption{CCDF of the empirical RMSE of the source-location estimation and its corresponding CRLB vs the outage threshold $\gamma$ for different values of $R_\mathsf{T}$ and $K_\mathsf{T}$.
	The observation SNR and channel SNR are $\mu = 40 \ \mathrm{dB}$ and $\eta = 0 \ \mathrm{dB}$, respectively.}
	\label{Fig:CCDF_GLE_NoSEx_KT_RandNet}
\end{figure}
}

\section{Conclusions}
\label{Sec:Conclusions}
The main focus of this paper was to quantify the effects of spatial randomness on the performance of source-localization schemes. To this end, a recently proposed approach based on the quantized versions of the received energies from a point source was investigated for demonstration purposes.
The random realization of the network geometry was assumed to be according to a uniform clustering process. The concept of localization outage was defined to be a realization of the network geometry that on average fails to satisfy a required threshold on the localization accuracy. The numerical results verified that the source-localization performance is heavily affected by the realization of sensor deployment and that it highly depends on the number of sensors that are within a close proximity of the source. This conclusion suggests a guideline that the sensor density in the network should appropriately be chosen such that enough number of sensors will be close to a target arbitrarily located within a random realization of the network geometry. As the network density increases, resulting in a higher number of sensors in a fixed disk around the source, the performance of the localization scheme improves drastically. The effect of exclusion zones around sensors was also studied based on which increasing the minimum sensor separation increases the localization-outage probability, i.e., if the sensors are forced to be farther separated, it is more likely that a random network realization will be in outage.

\ifCLASSOPTIONcaptionsoff
  \newpage
\fi


\fontsize{9.5}{12}
\selectfont
\bibliographystyle{IEEEtran}
\bibliography{Refs}

\end{document}